\documentclass{sig-alternate-05-2015}

\graphicspath{{./figure/}} 
\usepackage{epstopdf}
\usepackage[inline]{enumitem}
\usepackage{subcaption}

\usepackage{array}
\newcolumntype{L}[1]{>{\raggedright\let\newline\\\arraybackslash\hspace{0pt}}m{#1}}
\newcolumntype{C}[1]{>{\centering\let\newline\\\arraybackslash\hspace{0pt}}m{#1}}
\newcolumntype{R}[1]{>{\raggedleft\let\newline\\\arraybackslash\hspace{0pt}}m{#1}}

\begin{document}






%

\title{Tracking Illicit Drug Dealing and Abuse on Instagram using Multimodal Analysis}

\numberofauthors{2} 
%
\author{
%
%
\alignauthor
Xitong Yang\\
       \affaddr{University of Rochester}\\
       \email{yangxitongbob@gmail.com}
\alignauthor
Jiebo Luo\\
       \affaddr{University of Rochester}\\
       \email{jluo@cs.rochester.edu}
}

\maketitle
\begin{abstract}
Illicit drug trade via social media sites, especially photo-oriented Instagram, has become a severe problem in recent years. As a result, tracking drug dealing and abuse on Instagram is of interest to law enforcement agencies and public health agencies. In this paper, we propose a novel approach to detecting drug abuse and dealing automatically by utilizing multimodal data on social media. This approach also enables us to identify drug-related posts and analyze the behavior patterns of drug-related user accounts. 
To better utilize multimodal data on social media, multimodal analysis methods including multitask learning and decision-level fusion are employed in our framework. Experiment results on expertly labeled data have demonstrated the effectiveness of our approach, as well as its scalability and reproducibility over labor-intensive conventional approaches.
\end{abstract}

%
%


%
%

%
%



\begin{figure}
    \centering
    \includegraphics[width=.9\columnwidth]{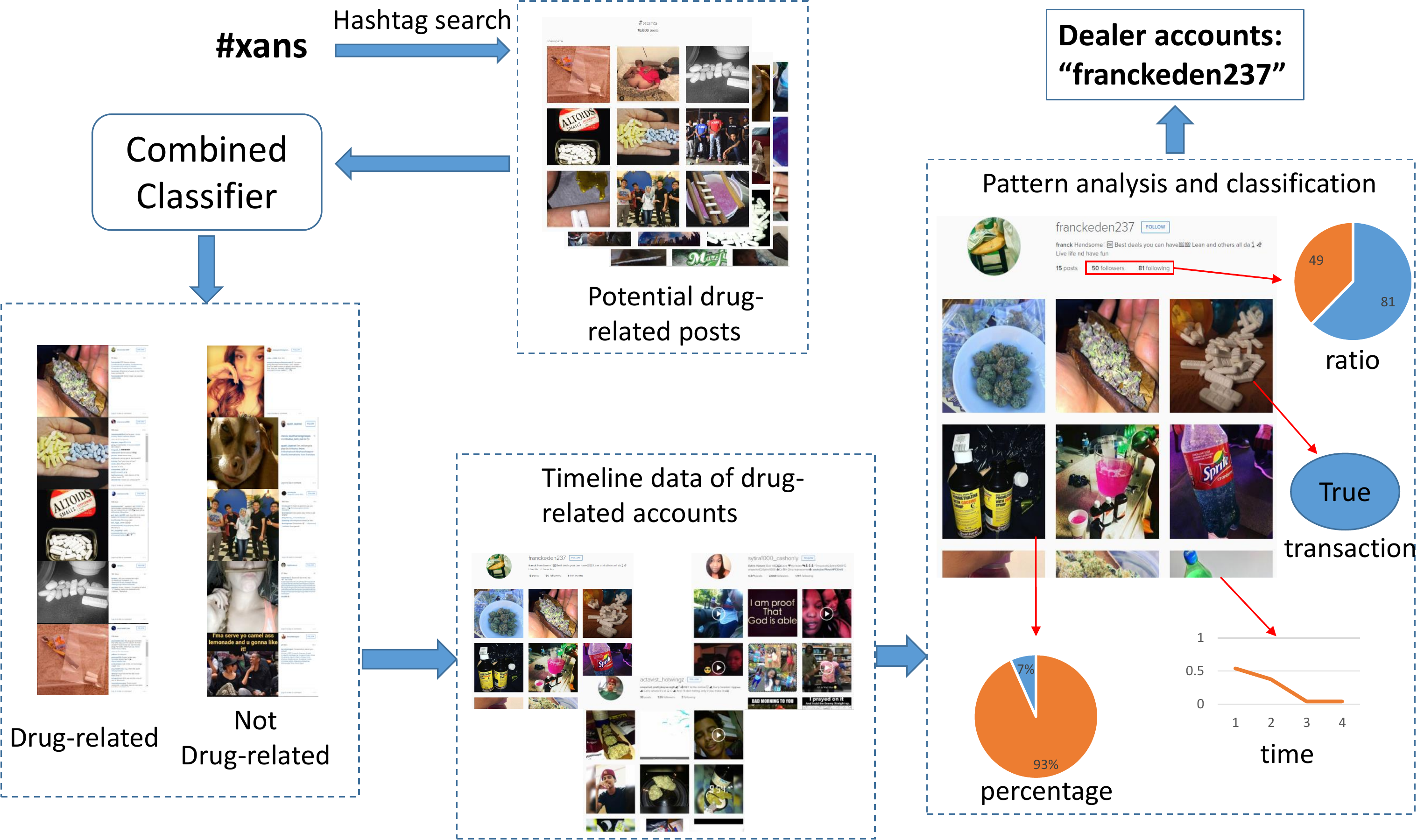}
    \caption{Exemplary illustration of our proposed approach to detecting drug dealer accounts. ``Percentage'' indicates the percentage of drug-related posts (orange). ``Ratio'' indicates the ratio of the number of followers (orange) and followees (blue). ``Time'' indicates the frequencies of drug-related posts in hours of a day. ``Transaction'' indicates the evidence of transaction.}
    \label{fig:example}
\end{figure}

\section{Introduction}
\label{sec:intro}
Nowadays, illicit drug use has become a major problem for the society and the trend has been steadily growing. According to the statistics of the annual National Survey on Drug Use and Health (NSDUH) \footnote{https://www.drugabuse.gov/publications/drugfacts/nationwide-trends}, an estimated 24.6 million Americans aged 12 or older (9.4 \% of the population) had used an illicit drug in the past month in 2013. This number is up from 8.3 \% in 2002. 
Meanwhile, with a massive number of active users, social media sites have become incredibly effective tools for advertising illegal drugs. For example, thousands of accounts on Instagram are currently selling marijuana, prescription painkillers, and other illicit drugs effectively in an open drug market. As a result, one can easily get access to these illegal drugs by a quick hashtag search on Instagram. It has been noticed and reported by law enforcement agencies an public health agencies that illegal drug trade on social media will aggravate drug abuse, especially for teenagers \footnote{http://www.news4jax.com/news/local/teens-getting-drugs-through-instagram} \footnote{http://wordondastreet.com/feds-use-instagram-arrest-350-drug-dealers-seize-7-million-one-weekend/}. Therefore, it is imperative to tackle drug dealing and abuse on social media.

Due to the anonymity on Instagram, drug dealers always post their offers in the most blatant fashion, which also provides us a good opportunity to track them through the contents they posted. However, traditional approaches, such as manual search and browsing by trained domain experts, suffer from two major problems when dealing with social media data. (1) \textit{Scalability}: it is impractical and inefficient to check the large-scale social media data manually. (2) \textit{Reproducibility}: even domain experts may overlook the evidences of drug dealing in some cases due to human errors. Moreover, different people may have different criteria for drug dealers, which makes the result inconsistent and not reproducible. To alleviate these problems, we propose to use machine learning algorithms to detect drug dealer accounts, which offers an effective and efficient way to assist the tracking of illicit drug trade. Note that some effort has been made recently to mine social media to track public health \cite{venkata2015social}, alcohol consumption \cite{pang2015monitoring} and even drug use patterns \cite{yiheng2016understanding, correia2016monitoring}.

\begin{figure}
    \centering
    \includegraphics[width=.9\columnwidth]{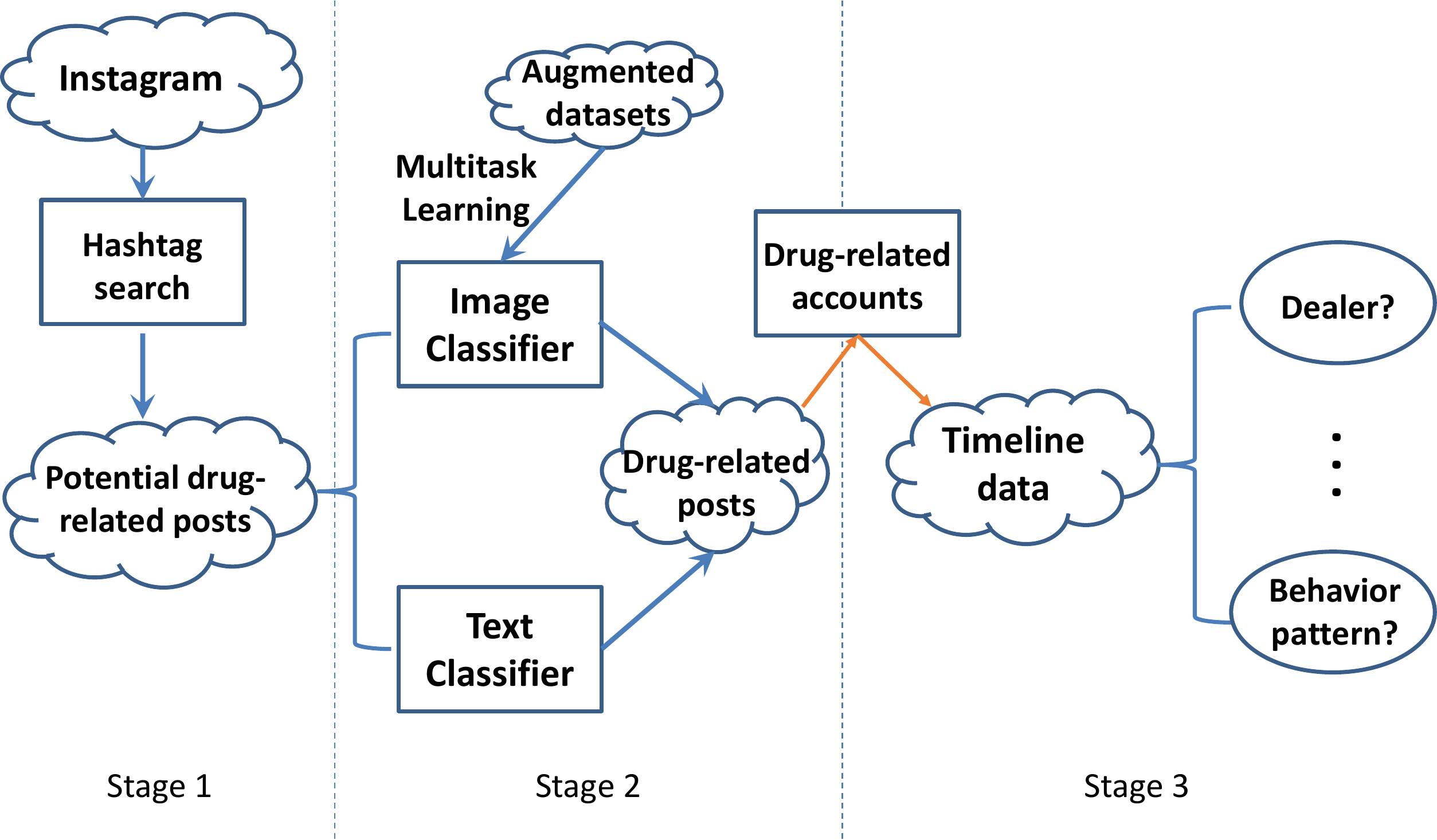}
    \caption{The framework of our proposed approach.}
    \label{fig:framework}
\end{figure}

In this paper, we propose a scheme to identify drug-related posts, analyze behavior patterns and finally detect drug dealer accounts. Multimodal data such as image, text, relational information and temporal pattern are used at different stages to improve the performance of the overall system. Specifically, the procedure includes the following three stages. First, given a dictionary of terms related to drug dealing provided by domain experts, we collect a pool of potential drug-related posts by a hashtag-based search on Instagram. Next, to identify real drug-related posts, image-based and text-based classifiers are both trained and used to filter these potential posts. A multitask learning method is employed to improve the image-based classifier using augmented datasets collected through a search engine. An intermediate decision is made by integrating the decisions of the two classifiers and corresponding drug-related accounts are identified. Finally, the timeline data of the drug-related accounts are collected before we analyze various activity and behavior patterns of these accounts to determine drug dealer accounts based on a set of selected features. The entire procedure is shown in Figure \ref{fig:framework}.

Using the example in Figure \ref{fig:example}, we further illustrate our approach to detect drug dealer accounts in a real case. First, we randomly pick a term in the dictionary (``xans" in this example) and use it as the keyword to search on Instagram. For simplicity, we download the 30 most recent posts as an example. Second, using our trained image-based and text-based classifiers, we identify 18 of them as drug-related posts, including 11 unique accounts (5 posts of each class are shown). Finally, we collect the timeline data of these 11 drug-related accounts, extract their behavior patterns as features and apply our trained drug dealer classifier on them. As a result, two accounts (``franckeden237'' and ``hustleboy420'') are detected as drug dealer accounts and we can prove the correctness by manually inspecting their pages. This example also shows the effectiveness and efficiency of our approach: using only 1 hashtag as the search keyword and 30 search outcomes, our approach can effectively discover two drug dealers on Instagram fully automatically.

Since hashtag search in stage 1 is trivial once we build a reliable dictionary, we focus on the last two stages in this paper. We will first describe how to train robust classifiers for drug-related post recognition, then move on to account pattern analysis and dealer account classification. The main contributions of this paper are as follows. 
\begin{itemize}
\item We propose an effective multimodal approach to tracking drug dealing and abuse on Instagram. 
\item We propose a multitask learning method to leverage web images for recognition tasks on social media. 
\item Our approach is generalizable to other similar problems, such as human trafficking and illegal gun sale.
\end{itemize}


\section{Drug-related posts recognition}
In this section, we describe how to train the classifiers for drug-related post recognition. Specifically, we first train the image-based and text-based classifiers separately. Decision-level fusion is then used to integrate two decisions as a weighted average. 

\subsection{Multitask learning for improved image-based classifier}
Image-based classifier is essential to identify drug-related posts. However, there are two major challenges when training a robust classifier. First, the appearances of drug-related posts are very diverse because of the different kinds of drugs and the intrinsic noise of social media posts. Second, the amount of labeled data is limited as manual annotation for images is tedious and time-consuming, which makes it more difficult for a classifier trained on limited data to generalize well to diverse cases. As a result, data augmentation is necessary to build a robust classifier.

Web data provides us a good option to build the augmented dataset. Specifically, we collect data from an image search engine (Google Image Search) by searching the hashtags in the aforementioned dictionary as keywords. The outcomes can be viewed as data with noisy labels because they are representative images for the search keywords. Using image search engine-based data as representative images is also adopted in \cite{li2015semantic,chen2015webly}, which did not use the data for multitask learning. The data is then grouped into three predefined sub-classes based on their appearances: weed, pills and syrup. These three sub-classes cover the most common illicit drugs, as reported in \cite{yiheng2016understanding}.

However, it is nontrivial to combine the original and augmented datasets coherently for training. The first challenge is the different data domains: while one comes purely from Instagram, the other comes from diverse web data. The second challenge is the different annotation settings, which means the labels of different datasets have different meanings (drug, weed, pills, syrup). Therefore, we need to transfer the knowledge about both data and task from the source domain (web dataset) to the target domain (Instagram dataset). 

\begin{figure}[t]
    \centering
    \begin{subfigure}[b]{.475\columnwidth}
        \includegraphics[width=.95\textwidth]{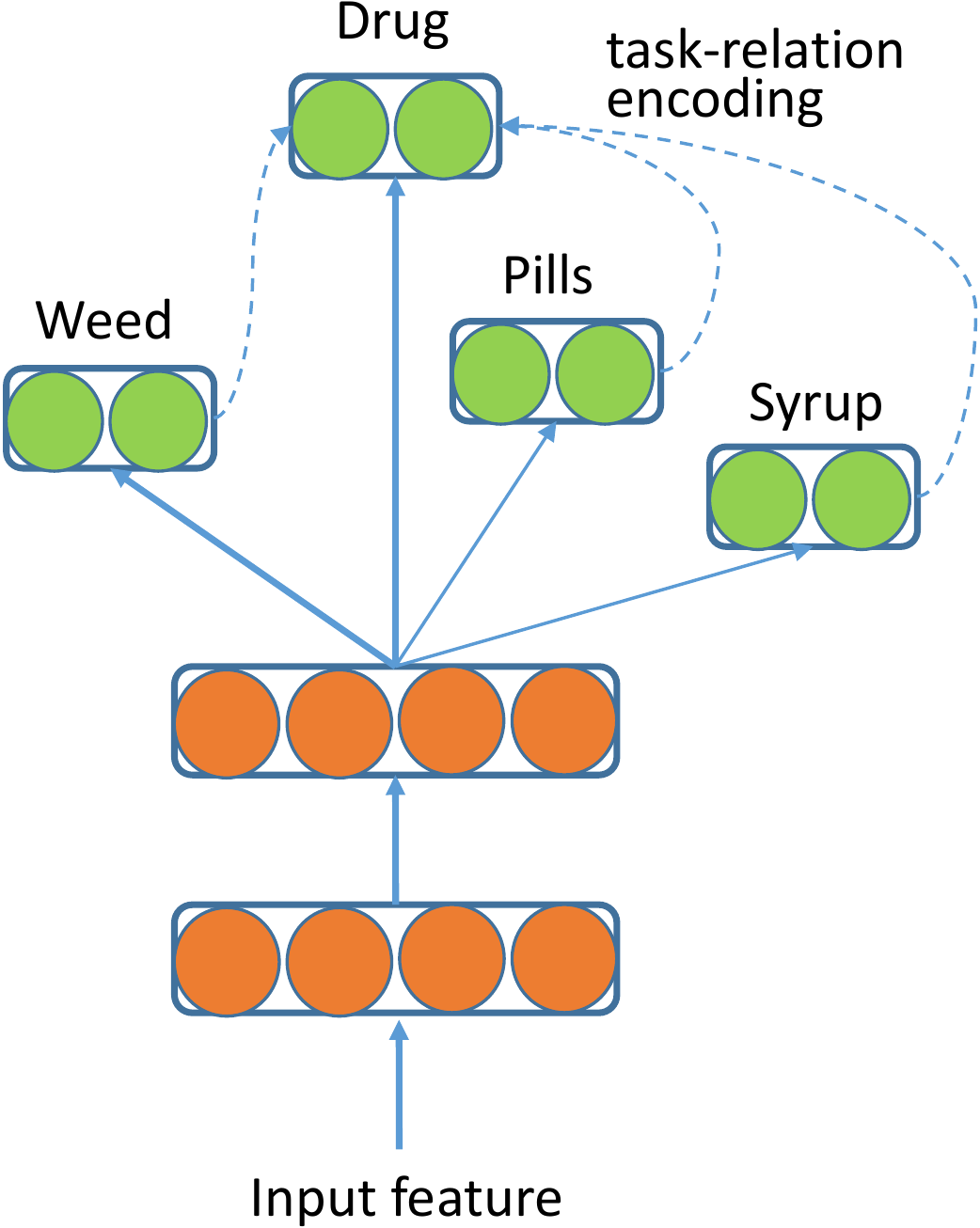}
        \caption{}
        \label{fig:model}
    \end{subfigure}
    ~
    \begin{subfigure}[b]{.475\columnwidth}
        \includegraphics[width=.95\textwidth]{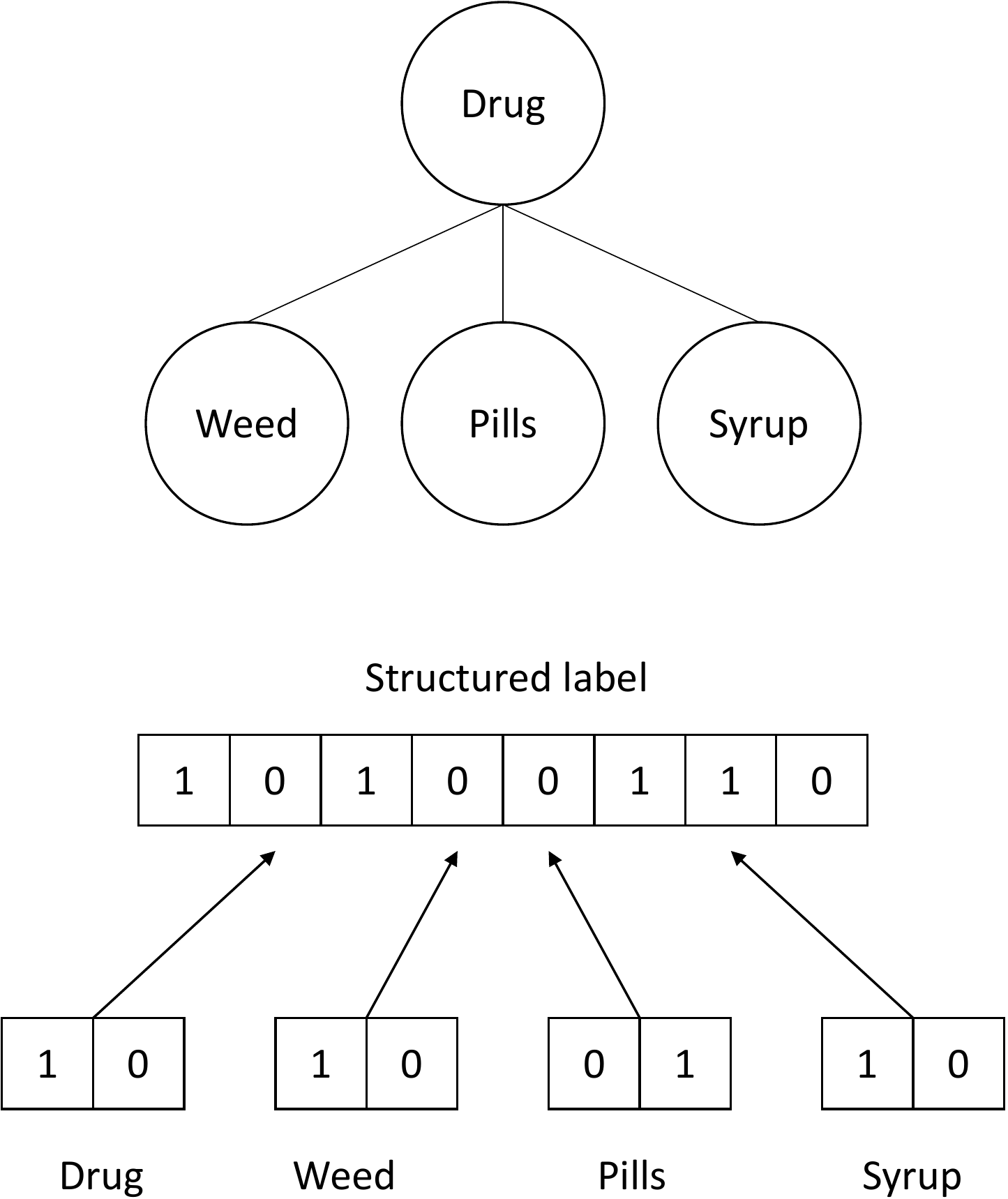}
        \caption{}
        \label{fig:label}
    \end{subfigure}
    \caption{(a) Model architecture for multitask learning. Dash lines indicates ``task-relation encoding". (b) Top: Task relationship can be encoded into a tree structure. Bottom: Structured representation of multitask predictions.}
        \label{fig:multitask}
\end{figure}

We propose a simple multitask learning method to solve this inductive transfer learning problem by learning common representations for relevant tasks \cite{collobert2008unified}. Inspire by \cite{wang2015your}, we first combine different label settings into a unified structured label representation, as illustrated in Figure \ref{fig:label} (Bottom). Moreover, task relationship is encoded into the structure representation. In our case, the four tasks can be represented as a tree structure as in Figure \ref{fig:label} (Top), where a parent node is a super-concept of its children nodes. As a result, we activate the parent node (labeled as positive) if one of its children nodes is activated. We call it \textit{task-relation encoding} for future reference. Intuitively, it means that whenever we see a ``weed" image, we should also classify it as a drug-related image. The model architecture is based on a multi-layer neural network, in which the lower level layers are shared among different tasks and the top-most layers are independent for different predictions. We illustrate the architecture in Figure \ref{fig:model}. The objective function is 
\begin{equation}
L = -\frac{1}{N} \sum_{n=1}^N \sum_{t \in T} \left[y_{nt}\log \hat{y}_{nt} + (1-y_{nt})\log (1-\hat{y}_{nt}) \right]
\end{equation}
where $T$ indicates the set of tasks we consider for data $n$.

Several details are worthnoting when optimizing the model. First, a masking mechanism is applied during training to ignore the irrelevant losses. Formally, as different nodes without direct connections should be mutually exclusive, we mask out the losses of all the unconnected nodes, given that the input data falls in a specific node. For example, when data of ``weed" dataset is fed to the model, we ignore the losses calculated by the predictors of ``pills" and ``syrup". In addition, although task-relation encoding introduces the dependency of tasks, this information is relatively noisy. Thus we set a lower weight (0.5) for the loss that comes from task-relation encoding prediction.

\subsection{Text-based classifier}
To train a text-based classifier, we first extract different kinds of features for tags and captions. Specifically, we extract uni-gram feature only for tags and both uni-gram and bi-gram features for captions. Features are then scaled by tf-idf weighting \cite{Salton:1986:IMI:576628}. To reduce the feature dimension, we retain the top 1000 features ordered by term frequency across the corpus for tags and comments, respectively. As a result, we have a 2000 dimensional vector for each post. A naive Bayes classifier \cite{mccallum1998comparison} is used for text classification.

\subsection{Evaluation}
We evaluate the performance of post classification at this stage by training and testing on a sample dataset we collect. Through hashtag-based search on Instagram,  we collect and manually annotate 4819 posts. Among these posts, only 1260 posts are annotated as positive, indicating that tags filtering alone is not robust enough. Using Google Image Search, we collect 4329 images in total and group them into weed, pills and syrup, in the number of 675, 2421 and 1233, respectively. Negative samples are obtained from data randomly collected on Instagram. Sample Instagram images and web images are shown in Figure \ref{fig:sample}.

For the image-based classifier, we first extract visual feature of each image using GoogLeNet \cite{szegedy2015going}. The features are then fed to the MTL framework as described above. We use two layers of shared hidden layers with the size of 256. Four softmax layers are connected for class prediction. The model is trained using gradient descent with RMSprop update rules. Both classifiers predict the probabilities of a post to be drug-related. A linear weighting method is used for decision-level fusion and we empirically pick the weight as 50\% for the image-based classifier and 50\% for the text-based classifier.

Five-fold cross validation is conducted to test the performance of the proposed methods. Different metrics: accuracy, precision, recall, f1 score and area under the curve (AUC) are reported in Table \ref{tab:result}. The best performance is achieved by fusing two classifiers.

\begin{figure}[t]
    \centering
    \begin{subfigure}[b]{.475\columnwidth}
        \includegraphics[width=1.0\textwidth]{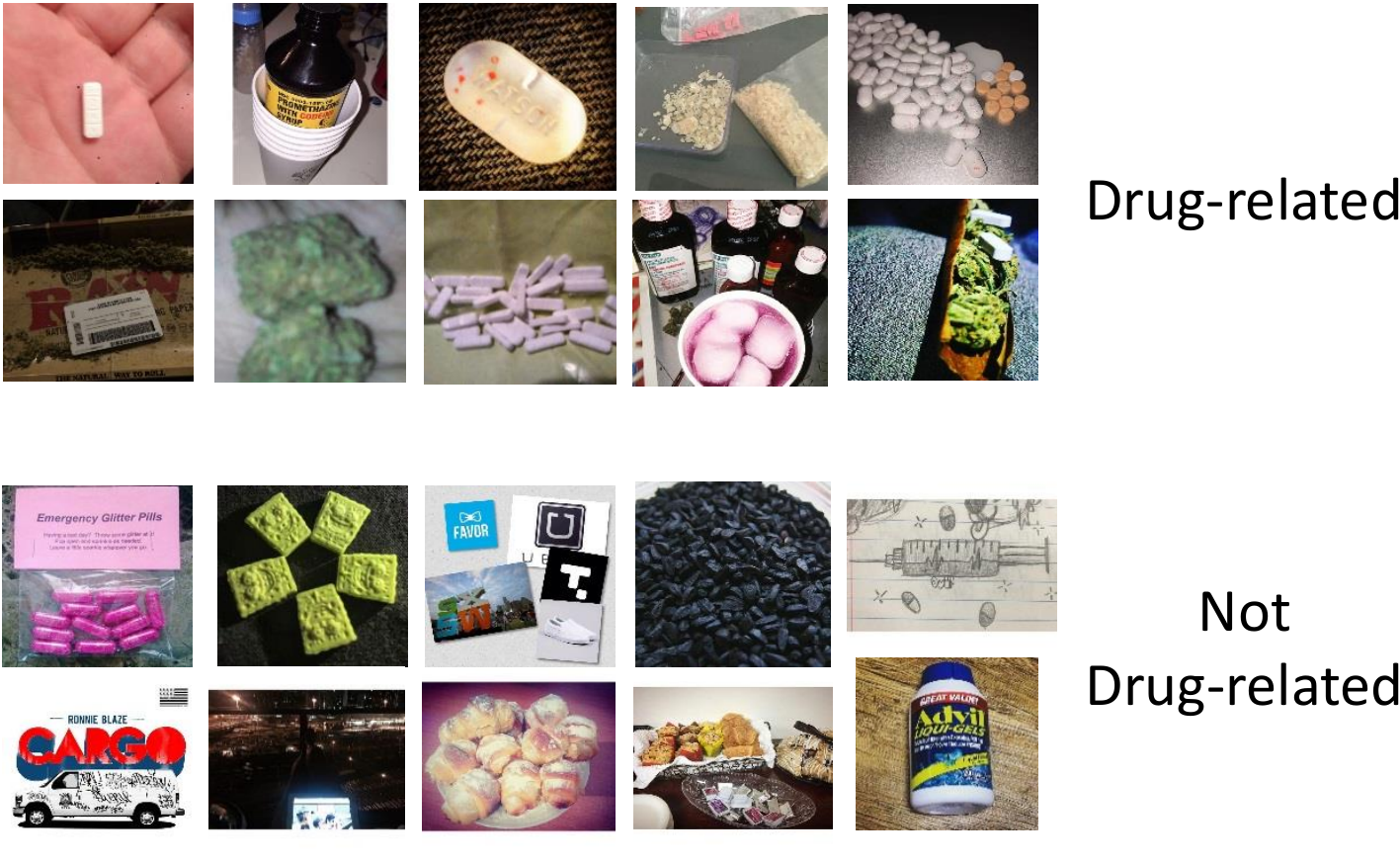}
        \caption{}
        \label{fig:sample}
    \end{subfigure}
    ~
    \begin{subfigure}[b]{.475\columnwidth}
        \includegraphics[width=1.0\textwidth]{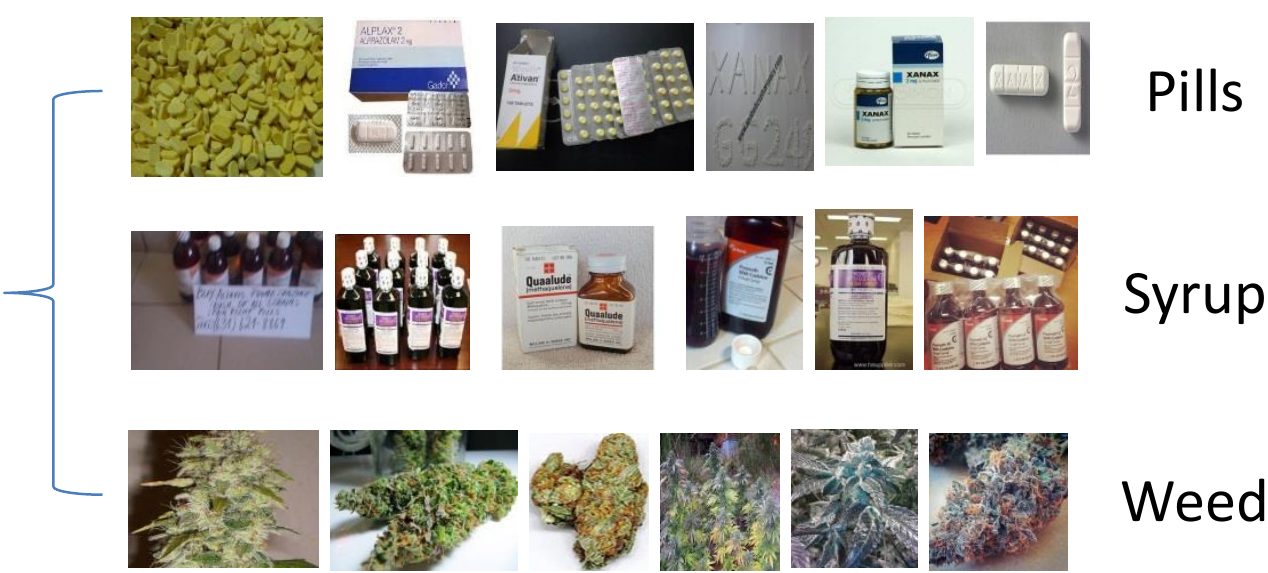}
        \caption{}
        \label{fig:sample2}
    \end{subfigure}
        \caption{Sample (a) Instagram images and (b) search engine-based images}
    \label{fig:sample}
\end{figure}

\section{Account Pattern Analysis}
In this section, we analyze several behavior patterns of drug-related accounts using their timeline data. We collect a sample dataset of 206 drug-related accounts and each of them has up to 200 recent posts for simplicity. After the annotation by domain experts, 27 of these accounts are labeled as drug dealers with evidences of drug transactions. We note that geolocation information is also useful for tracking drug dealers. However, as the location information of most posts are unavailable, we do not analyze the geolocation pattern of drug-related accounts.

\begin{table}
\centering
\caption{Results of drug-related post recognition.}
\begin{tabular}{c | c c c c}
Methods & Accuracy & Precision & Recall & F1-score\\ \hline
MLP & 86.9 & 80.3 & 66.3 & 0.72\\
MT\_MLP & 87.2 & 81.0 & 66.7 & 0.73 \\ \hline
Tags only & 82.2 & 67.2 & 62.0 & 0.65 \\
Caption only & 81.7 & 69.4 & 53.7 & 0.61 \\
Combined & 81.7 & 63.6 & 70 & 0.67 \\ \hline
Late fusion & \textbf{88.1} & \textbf{83.1} & \textbf{68.1} & \textbf{0.75}\\
\end{tabular}
\label{tab:result}
\end{table}

\subsection{Percentage of drug-related posts}
With a trained classifier for drug-related posts, it is straightforward to obtain the percentage of drug-related posts within a user's timeline data, which is calculated by (\# of drug-related posts / \# of all posts). The value of the ratio can be used as a metric to distinguish drug dealer accounts on Instagram. Figure \ref{fig:ratio} shows the histogram of the percentages of dealer accounts and non-dealer accounts. It is clear that accounts with a higher value of drug-related percentage are more likely to be drug dealers, while most of the non-dealer accounts have a very low percentage of drug-related posts.

\subsection{Temporal patterns}
Temporal patterns refer to the frequencies of drug-related posts along the time axis. We analyze the temporal patterns in different timescales, and find that the signal is most significant when we focus on different hours of a day. The statistics are shown in Figure \ref{fig:time}. We can see that drug dealers tend to post more drug-related posts at midnight.

\subsection{Relational information}
Relational information is also useful to distinguish drug dealer accounts. As expected, drug dealer accounts should have more followers than their following accounts, as they are selling their products to their followers. That means the ratio (\# of following / \# of follower) is quite low for them. Our analysis confirms this intuition in some sense, as shown in Figure \ref{fig:follow}.

\subsection{Evidences of transactions}
\begin{figure}[t]
    \centering
\begin{subfigure}[t]{.475\columnwidth}
    \includegraphics[width=.9\textwidth]{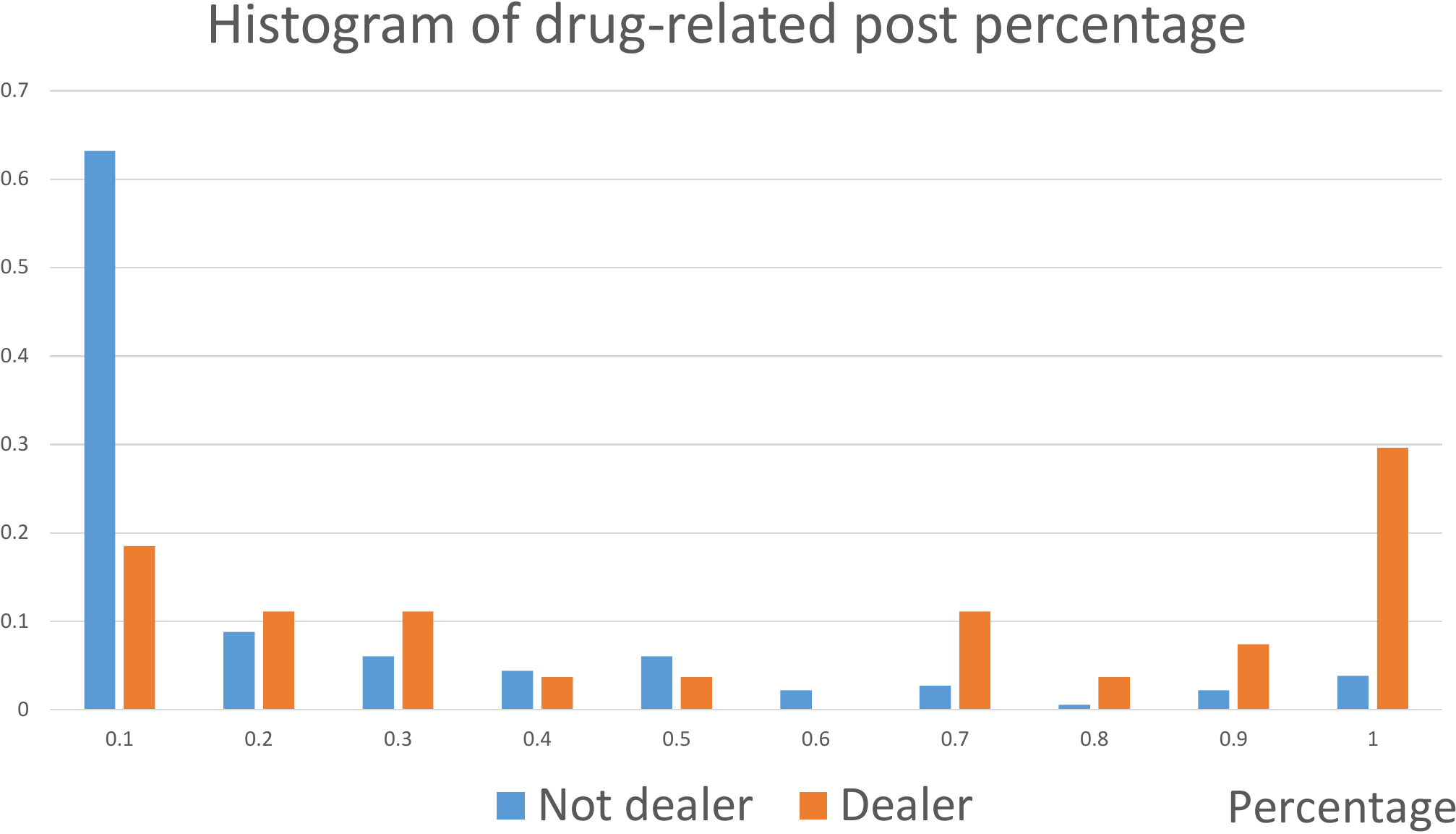}
    \caption{}
    \label{fig:ratio}
\end{subfigure}
~
\begin{subfigure}[t]{.475\columnwidth}
    \includegraphics[width=.9\textwidth]{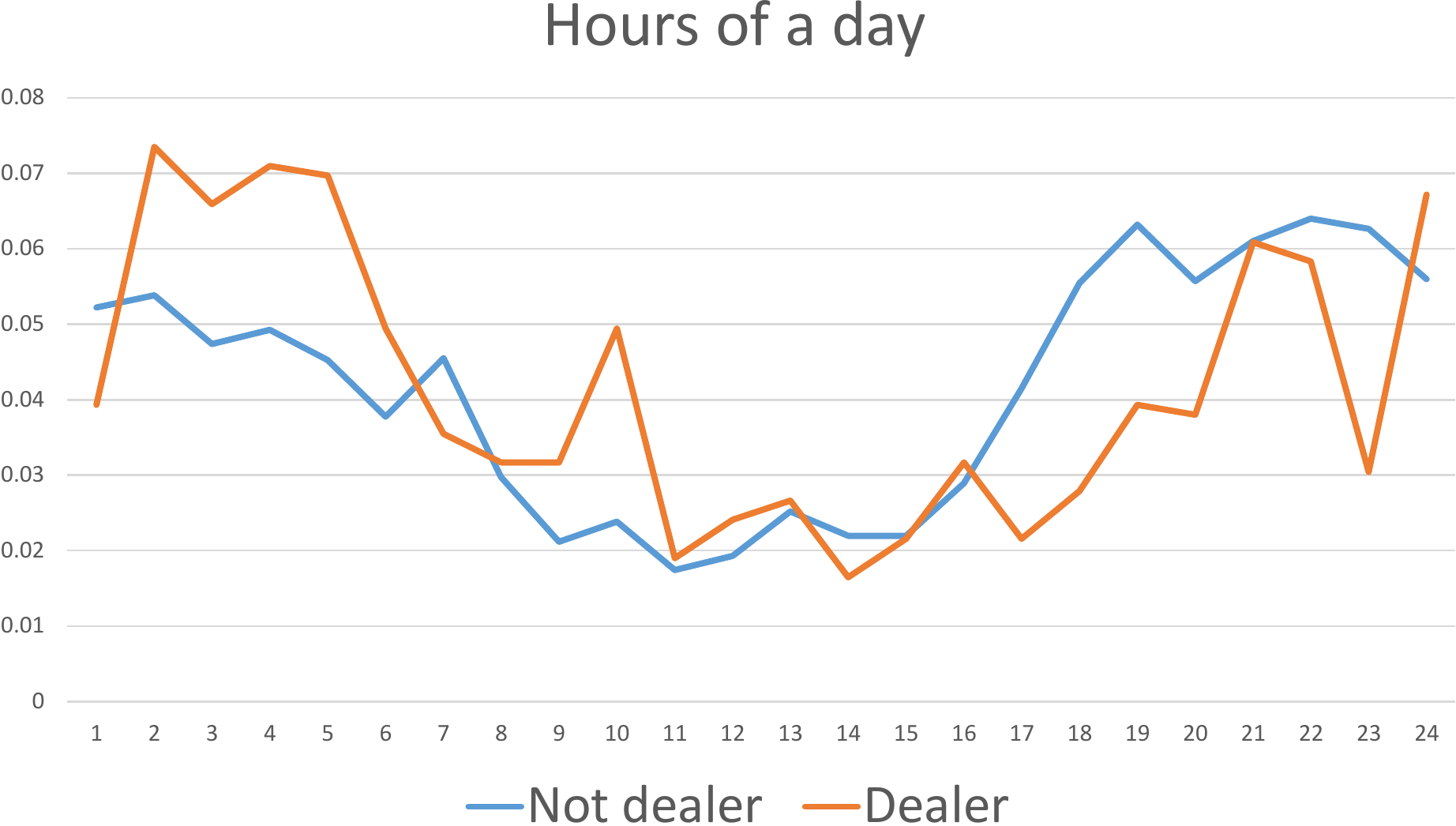}
    \caption{}
    \label{fig:time}
\end{subfigure}
~
\begin{subfigure}[b]{.475\columnwidth}
    \includegraphics[width=.9\textwidth]{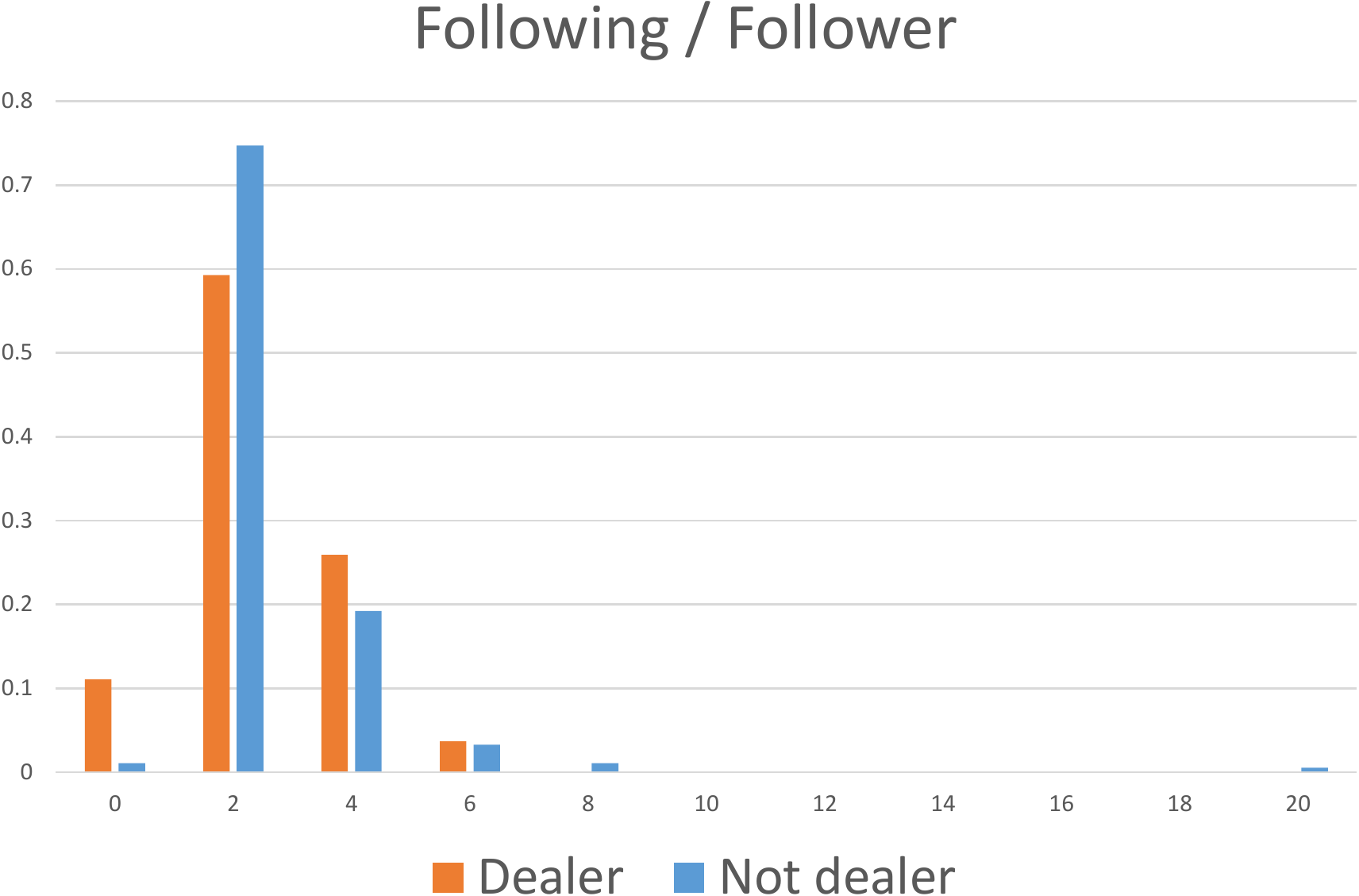}
    \caption{}
    \label{fig:follow}
\end{subfigure}
~
\begin{subfigure}[b]{.475\columnwidth}
    \includegraphics[width=.9\textwidth]{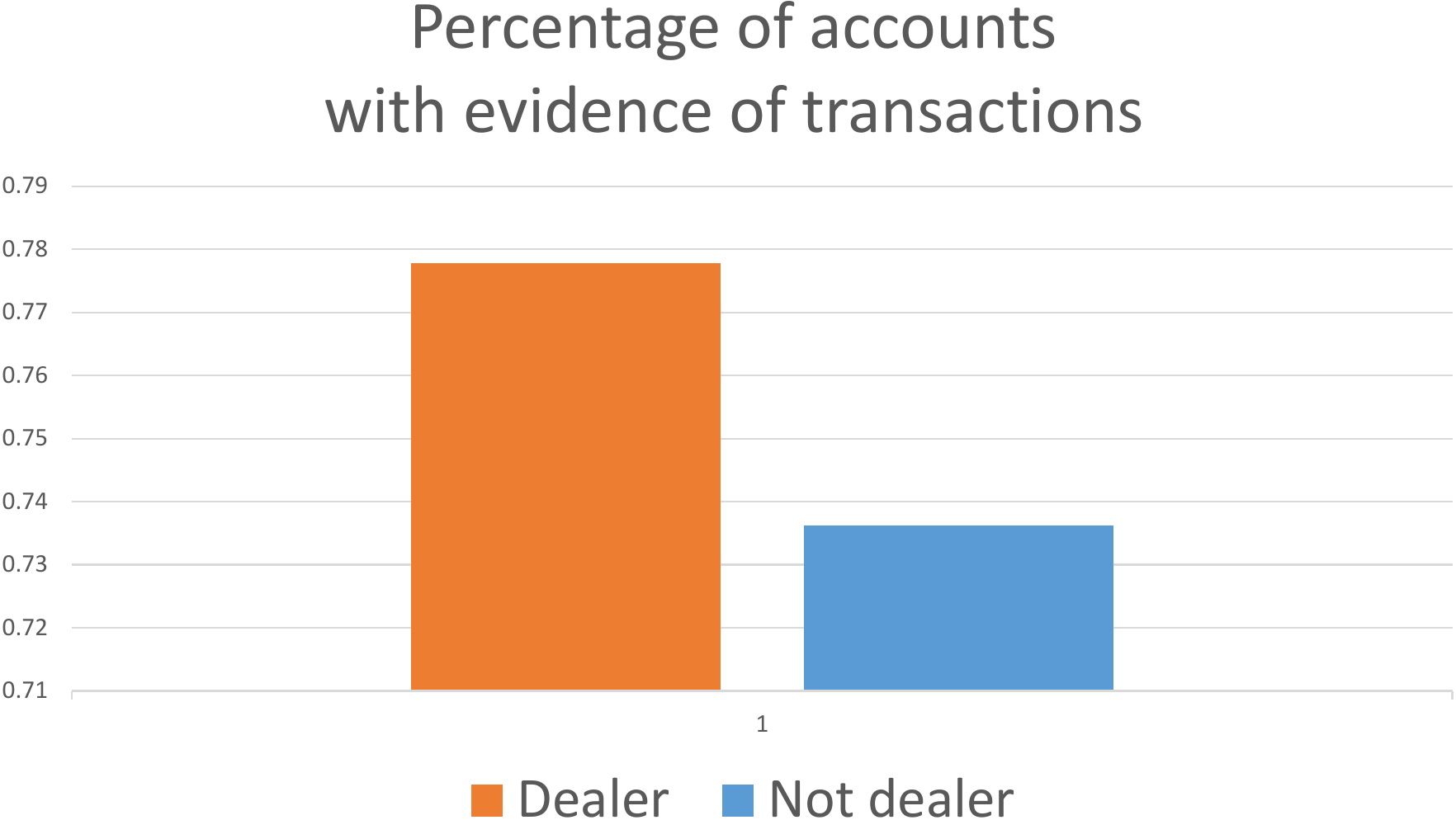}
    \caption{}
    \label{fig:transaction}
\end{subfigure}

\caption{(a) Drug-related post ratio of dealers and non-dealers. (b) Temporal pattern of drug-related accounts. (c) Relational information pattern of drug-related accounts. (d)Percentage of accounts with evidence of transactions.}
\end{figure}

Evidences of transactions is another criterion to determine a drug dealer account. As stated by the domain experts, one account will not be identified as drug dealer account if it does not contain evidences of drug transactions, even though it contains many drug-related posts. This inspires us to capture the evidences of transactions in the timeline data. Specifically, we extract the bio description and comments of the drug-related posts. A predefined blacklist provided by the domain experts are applied to filtering the content. Once the terms of blacklist occur, we assign a 'True' value to this feature, otherwise we assign a 'False' value. Figure \ref{fig:transaction} shows the percentage of accounts with evidence of transactions. Even though our method cannot detect all the transaction information, we still can see the trend that dealer accounts have a higher percentage value than non-dealer accounts.

\section{Dealer accounts detection}
Finally, we use the aforementioned modalities with significant signals as features to detect drug dealer accounts. Specifically, we use the following features as input:
\begin{itemize}
\item $P$: Percentage of drug-related post (1D)
\item $H_1$ to $H_4$: Hour of a day (binned to 4D)
\item $R_1$ to $R_3$: Relational information: \# of Follower, \# of Following, Following/Follower (3D)
\item $E$: Evidence of transaction (1D)
\end{itemize}

Features are normalized to zero mean and unit variance.

\subsection{Feature selection}
Although the above features show useful signals in pattern analysis, we are still not sure about their significance for detecting drug dealer accounts. We choose a feature selection approach using L1 regularization. In particular, we train a linear logistic regression classifier with L1 regularization on hold-out data and remove the feature dimensions with zero or very small coefficient values. Five features are retained after the feature selection, namely $P$, $H_1$ (midnight), $H_4$ (late night), $R_1$ (\# of follower), $R_3$ and $E$. Five-fold cross validation is then performed to evaluate the method and the result is showed in Table \ref{tab:dealer}.

\subsection{Inconsistency in human annotation}
As we mentioned in Section \ref{sec:intro}, even domain experts may overlook some evidences or have inconsistent judgments during annotation. In this experiment, we have two independent sets of labels annotated by two experts. 94 \% of the labels are consistent between the two experts. However, such seemingly minor differences resulted in very different evaluation scores as our dataset is highly imbalanced (significantly fewer positives).  Compared with annotation by human experts, our machine learning-based approach offers reproducible and consistent prediction. Table \ref{tab:dealer} supports this observation.

\begin{table}[h!]
\centering
\begin{tabular}{c c|c c c }
Method & Expert ID&Precision & Recall & F1-score \\ \hline
LR & exp1 & 28.8 & 55.6 & 0.38\\
LR & exp2 & 38.5 & 60.6 & 0.47 \\ \hline
LR-select & exp1 & 30.6 & 55.6 & 0.40\\
LR-select & exp2 & 42.9 & 63.6 & 0.51\\
\end{tabular}
\caption{Results of drug dealer accounts classification with two experts. We regard each expert's annotation as the ground truth to obtain the evaluation scores for the other expert.}
\label{tab:dealer}
\end{table}

\section{Conclusions}
In this paper, we propose a comprehensive framework for identifying drug-related posts, analyzing behavior patterns and detecting drug dealer accounts on Instagram. Multimodal data and analysis methods are employed to achieve high quality results competitive with domain experts. The experiments show that our proposed approach is efficient and reproducible for practical use, which can serve as an effective tool for tracking and combating illicit drug trade on social media. 

\section{Acknowledgment}
This work was supported in part by New York State through the Goergen Institute for Data Science at the University of Rochester. We thank Meredith McCarran, Lacey Keller and Ruth Hirsch of the NYSAGO for the valuable assistance to this work.

\section{Disclaimer}
This report does not reflect the views, opinions or endorsement of the NYOAG, unless the NYOAG specifically waives this requirement.

%
\newpage
\bibliographystyle{abbrv}
\bibliography{Drug}  
%

\end{document}